\documentclass[acus]{JAC2003}

\usepackage{graphicx}
\usepackage{amsmath}

\begin{document}

\title{STRUCTURE OF THE NUCLEON FROM ELECTROMAGNETIC SPACE-LIKE AND TIME-LIKE FORM
FACTORS}
\author{F. Iachello\\
Center for Theoretical Physics, Sloane Physics Laboratory,\\
Yale University, New Haven, Connecticut 06520-8120, USA}
\maketitle

\begin{abstract}
Recent experimental data on space-like and time-like form factors of the
nucleon are reviewed in light of a model of the nucleon with an intrinsic
(quark-like) structure and a meson cloud. The analysis points to the
astonishing result that the proton \textit{electric} space-like form factor 
\textit{vanishes} at $Q^{2}\sim 8$ (GeV/c)$^{2}$ and becomes negative beyond
that point. The intrinsic structure is estimated to have a r.m.s. radius of $%
\sim 0.34$ fm, much smaller than the proton r.m.s. radius $\sim 0.82$ fm.
The calculations are in perfect agreement with space-like proton data, but
deviate drastically from space-like neutron data at $Q^{2}>1$ (GeV/c)$^{2}$.
Analysis of the time-like data appears to indicate excellent agreement with
both proton and neutron data in the entire range of measured $q^{2}=-Q^{2}$
values.
\end{abstract}

\section{INTRODUCTION}

Electromagnetic form factors have played a crucial role in understanding the
structure of composite particles. A particularly important composite
particle is the nucleon, which forms the basis upon which all matter is
built. Studies of the structure of the nucleon with electromagnetic probes
begun in the late 50's and early 60's when Hofstadter and collaborators
demonstrated that the nucleon was not point-like with a (proton) root-mean
square radius $\langle r_{p}^{2}\rangle ^{1/2}\sim 0.75$ fm. In the 1970's
many experiments were performed, showing that the neutron was a complex
particle with a negative r.m.s. radius and $dG_{E_{n}}/d(Q^{2})\sim 0.50$
(GeV/c)$^{2}.$ In 1973, it was suggested that the nucleon has a two
component structure with an intrinsic part with form factor $g(Q^{2})$ and a
meson cloud parametrized in terms of vector mesons, $(\rho ,\omega ,\varphi
) $. In the late 1970's the non-relativistic quark-model was used to
describe the properties of hadrons. It was soon realized that this model
cannot describe form factors in a consistent way. Also in the late 1970's,
QCD emerged as the theory of strong interactions. In a perturbative
approach, p-QCD, the asymptotic behavior of the form factors can be derived,
yielding the large $Q^{2}$ behavior of the nucleon form factors to be $%
\propto \frac{1}{Q^{4}}$. \ In the 1980's, experimental groups noted that
all form factors, except $G_{E_{n}}$, could be described by the empirical
dipole form $G_{D}(Q^{2})=1/\left( 1+\frac{Q^{2}}{0.71}\right) ^{2}.$ These
observations culminated in the SLAC experiment NE11 on the ratio $\mu
_{p}G_{E_{p}}/G_{M_{p}}$ \ that appeared to be consistent with scaling up to 
$10$ (GeV/c)$^{2}$ \cite{andivahis}. However, in 2000-2002 experiments
performed at TJNAF \cite{jones}, \cite{gayou} using the recoil polarization
method have shown the astounding result that the ratio of proton electric to
proton magnetic form factor decreases dramatically with $Q^{2},$
inconsistent with scaling. In the first part of this article, the present
situation on electromagnetic form factors of the nucleon in the space-like
region will be reviewed.

For relativistic systems, one has access to the time-like part of the form
factors. Studies of this part begun in the late 60's and early 70's through
processes $p\bar{p}\rightarrow e^{+}e^{-}$ and $e^{+}e^{-}\rightarrow p\bar{p%
}$. The first positive result on the time-like form factors of the proton
was obtained in 1972 at\ Frascati \cite{castellano}. Several other
experiments where subsequently performed culminating in the Fermilab
experiment E835 \cite{ambrogiani}. In 1998 the first measurement of the
neutron time-like form factor was reported \cite{antonelli}. This
measurement appeared to be in disagreement with simple quark model
extensions to large $Q^{2}$, where $\mid G_{M_{n}}/G_{M_{p}}\mid =-2/3$. In
the second part of this article, the present situation on the time-like
electromagnetic form factors of the nucleon will be reviewed.

\section{SPACE-LIKE FORM FACTORS}

Two basic principles play a crucial role in the analysis of electromagnetic
form factors of the nucleon. The first of these is relativistic invariance.
This principle fixes the form of the nucleon current to be \cite{bjorken}
\begin{equation}
J^{\mu }=F_{1}(Q^{2})\gamma ^{\mu }+\frac{\kappa }{2M_{N}}%
F_{2}(Q^{2})i\sigma ^{\mu \nu }q_{\nu }
\end{equation}
where $F_{1}(Q^{2})$ and $F_{2}(Q^{2})$ are the so-called Dirac and Pauli
form factors and $\kappa $ is the anomalous magnetic moment. This symmetry
is expected to be exact. The second is isospin invariance. Although this
symmetry is not exact, being of dynamical origin, it is expected to be only
slightly broken in a realistic theory of strong interaction. Isospin
invariance leads to the introduction of isoscalar, $F_{1}^{S}$ and $%
F_{2}^{S} $, and isovector, $F_{1}^{V}$ and $F_{2}^{V}$, form factors, and
hence to relations among proton and neutron form factors. The observed Sachs
form factors, $G_{E}$ and $G_{M}$ can be obtained by the relations
\begin{eqnarray}
G_{M_{p}} &=&\left( F_{1}^{S}+F_{1}^{V}\right) +\left(
F_{2}^{S}+F_{2}^{V}\right)  \notag \\
G_{E_{p}} &=&\left( F_{1}^{S}+F_{1}^{V}\right) -\tau \left(
F_{2}^{S}+F_{2}^{V}\right)  \notag \\
G_{M_{n}} &=&\left( F_{1}^{S}-F_{1}^{V}\right) +\left(
F_{2}^{S}-F_{2}^{V}\right)  \notag \\
G_{E_{n}} &=&\left( F_{1}^{S}-F_{1}^{V}\right) -\tau \left(
F_{2}^{S}-F_{2}^{V}\right)
\end{eqnarray}
with $\tau =Q^{2}/4M_{N}^{2}$. These relations also satisfy another
constraint, namely the kinematical constraint $%
G_{E}(-4M_{N}^{2})=G_{M}(-4M_{N}^{2}).$ This constraint is of crucial
importance in the time-like region, while playing a minor role in the
space-like region.

Different models of the nucleon correspond to different assumptions for the
Dirac and Pauli form factors. In 1973 \cite{IJL} a model of the nucleon in
which the external photon couples to both an intrinsic structure, described
by the form factor $g(Q^{2})$, and a meson cloud, treated within the
framework of vector meson ($\rho ,\omega $ and $\varphi $) dominance, was
suggested. In this model the Dirac and Pauli form factors are parametrized as
\begin{eqnarray}
F_{1}^{S}(Q^{2}) &=&\frac{1}{2}g(Q^{2})[(1-\beta _{\omega }-\beta _{\varphi
})  \notag \\
&&+\beta _{\omega }\frac{m_{\omega }^{2}}{m_{\omega }^{2}+Q^{2}}+\beta
_{\varphi }\frac{m_{\varphi }^{2}}{m_{\varphi }^{2}+Q^{2}}] \\
F_{1}^{V}(Q^{2}) &=&\frac{1}{2}g(Q^{2})[(1-\beta _{\rho })+\beta _{\rho }%
\frac{m_{\rho }^{2}}{m_{\rho }^{2}+Q^{2}}]  \notag \\
F_{2}^{S}(Q^{2}) &=&\frac{1}{2}g(Q^{2})[\left( -0.120-\alpha _{\varphi
}\right) \frac{m_{\omega }^{2}}{m_{\omega }^{2}+Q^{2}}  \notag \\
&&+\alpha _{\varphi }\frac{m_{\varphi }^{2}}{m_{\varphi }^{2}+Q^{2}}] \\
F_{2}^{V}(Q^{2}) &=&\frac{1}{2}g(Q^{2})[3.706\frac{m_{\rho }^{2}}{m_{\rho
}^{2}+Q^{2}}]
\end{eqnarray}
In \cite{IJL} three forms of the intrinsic form factor $g(Q^{2})$ were used.
The best fit was obtained for $g(Q^{2})=(1+\gamma Q^{2})^{-2}$. This form
will be used in the remaining part of this talk. Before comparing with the
data, an additional modification is needed. In view of the fact that the $%
\rho $ meson has a non-negligible width, one needs to replace
\begin{equation}
\frac{m_{\rho }^{2}}{m_{\rho }^{2}+Q^{2}}\rightarrow \frac{m_{\rho
}^{2}+8\Gamma _{\rho }m_{\pi }/\pi }{m_{\rho }^{2}+Q^{2}+\left( 4m_{\pi
}^{2}+Q^{2}\right) \Gamma _{\rho }\alpha (Q^{2})/m_{\pi }}
\end{equation}
where
\begin{multline}
\alpha \left( Q^{2}\right) \\=\frac{2}{\pi }\left[ \frac{4m_{\pi }^{2}+Q^{2}}{%
Q^{2}}\right] ^{1/2}\ln \left( \frac{\sqrt{4m_{\pi }^{2}+Q^{2}}+\sqrt{Q^{2}}%
}{2m_{\pi }}\right) .
\end{multline}

This replacement is important for small $Q^{2}$, although, because of the
logarithm dependence of the $\pi \pi $ cut expressed by the function $\alpha
(Q^{2})$, its effect is felt even at moderate and large $Q^{2}$.

\subsection{The ratio of electric to magnetic form factors of the proton}

By using the coupling constants given in Table 1 of \cite{IJL} $\beta _{\rho
}=0.672,\beta _{\omega }=1.102,\beta _{\varphi }=0.112,\alpha _{\varphi
}=-0.052,$ an intrinsic form factor with $\gamma =0.25$ (GeV/c)$^{-2}$,
standard values of the masses ($m_{\rho }=0.765$ GeV$,m_{\omega }=0.784$ GeV$%
,m_{\varphi }=1.019$ GeV), and a $\rho $ width $\Gamma _{\rho }=0.112$ GeV,
one can calculate the ratio $\mu _{p}G_{E_{p}}/G_{M_{p}}$. The result is
shown against the new data \cite{jones}, \cite{gayou} in Fig.1. The
agreement is astonishing. Fig. 1 also shows the remarkable result that the 
\textit{electric} form factor of the proton crosses zero at $Q^{2}\sim 8$
(GeV/c)$^{2}$. It would be ot utmost importance to measure the ratio $\mu
_{p}G_{E_{p}}/G_{M_{p}}$ at $Q^{2}\geq 6$ (GeV/c)$^{2}.$ A measurement of
the zero of the electric form factor, adding to the already measured sharp
drop \ from $1$ at $Q^{2}=0$ to $\sim 0.27$ at $Q^{2}=5.6$ (GeV/c)$^{2}$,
would unequivocably establish the complex nature of the nucleon. In the
model put forward in 1973, the nucleon has both an intrinsic structure
(presumably three valence quarks) and additional contributions (presumably $q%
\bar{q}$ pairs). (The complex nature of the nucleon resulting from
electromagnetic form factors is in accord with results obtained by the $EMC$
collaboration \cite{EMC}, where the additional, non $q^{3}$, components were
attributed to gluons.) An estimate of the spatial extent of the intrinsic
region (where the fundamental quarks sit) can be obtained from the value of $%
\gamma $ in the intrinsic form factor. The r.m.s. of this distribution is $%
\sim 0.34$ fm, much smaller that the proton r.m.s. radius $\sim 0.87$ fm.
The zero in the \textit{electric }form factor is a consequence of the two
term structure of Eq.(2), in particular of the fact that the second term is
multiplied by $-Q^{2}/4M_{N}^{2}.$ Any model with a two term structure will
produce results in qualitative agreement with data. Indeed three of the
descriptions considered in \cite{gayou}, a soliton model \cite{holzwarth},
and two relativistic constituent quark models \cite{miller}, \cite{simula}
have this structure and produce results in qualitative agreement with
experiment. Also the introduction of relativity in non-relativistic quark
models goes in the direction of reducing the ratio \cite{desanctis}. To
discriminate between various models it is necessary to find precisely at
which value the zero occurs.

\subsection{The magnetic form factor of the proton}

The agreement between theory and data for the proton form factors is not
limited to the ratio $\mu _{p}G_{E_{p}}/G_{M_{p}}.$ Consider the magnetic
form factor, $G_{M_{p}}$. For convenience of display, normalize it to the
so-called dipole form factor, $G_{D}=(1+\frac{Q^{2}}{0.71})^{-2}$. The data 
\cite{bartel}, \cite{bosted}, \cite{andivahis} in the interval $0\leq
Q^{2}\leq 10$ (GeV/c)$^{2}$ are plotted in Fig.2. They show an ondulation,
crossing the value one at $Q^{2}\sim 0.6$ (GeV/c)$^{2}$ and again at $\sim 6$
(GeV/c)$^{2}$. The calculation is in excellent agreement with the data, with
crossing points at precisely the same values $\sim 0.6$ and $6$ (GeV/c)$^{2}$%
. The observed ondulation is proof that vector meson (with masses $\mu
^{2}\sim 0.5-1.0$ (GeV/c)$^{2}$) components are important. Without $\rho $
meson component, the form factor should behave smoothly (see Fig. 3 of \cite
{IJL}).%
\begin{figure}
\begin{center}
\includegraphics*[width=\hsize]{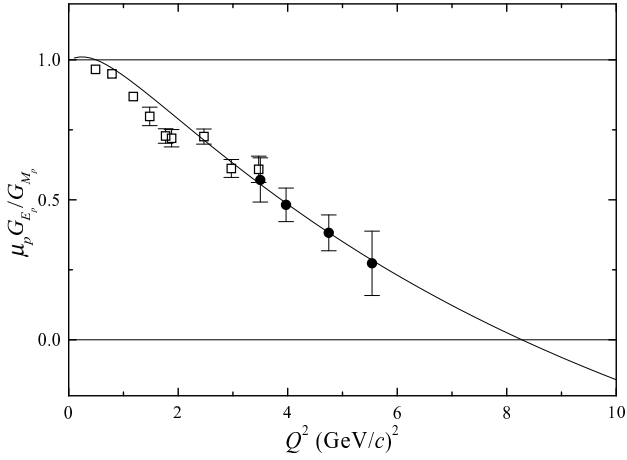}
\end{center}
\vspace{-12pt}
\caption{
The measured ratio $\mu _{p}G_{E_{p}}/G_{M_{p}}$ compared with the
1973 prediction. Ref. [2]: open square. Ref. [3]: filled circle.
}
\end{figure}
\begin{figure}
\begin{center}
\includegraphics*[width=\hsize]{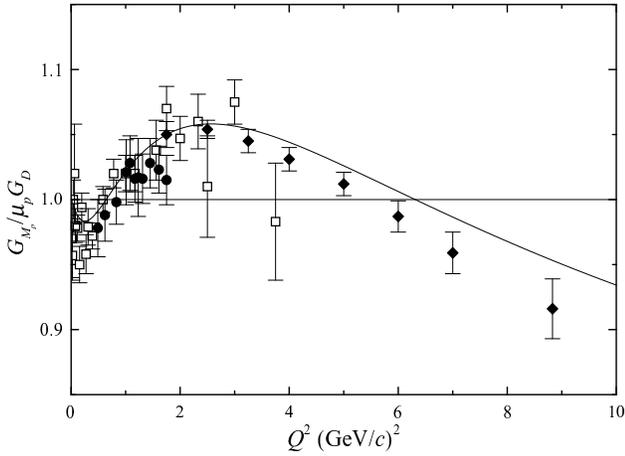}
\end{center}
\vspace{-12pt}
\caption{
Experimental values $G_{M_{p}}/\mu _{p}G_{D}$ compared with
calculation. Ref.[14]: open square. Ref.[15]: filled circle. Ref.[1]: filled
diamond.
}
\end{figure}

\subsection{The magnetic form factor of the neutron}

Having established the structure of the proton, I now come to that of the
neutron. This is dictated by isospin invariance. Measurements of the neutron
form factors are obscured by the knowledge of the wave functions of
deuterons or He$^{3}.$ Older measurements are either in disagreement (for $%
Q^{2}>1$ (GeV/c)$^{2}$) or in marginal agreement ($Q^{2}<1$ (GeV/c)$^{2}$)
with the 1973 model. However, the situation here appears to be similar to
the situation for the proton form factors previous to the experiments of
Jones et al \cite{jones} and Gayou et al \cite{gayou}. I consider first the
region $Q^{2}\leq 1$ (GeV/c)$^{2}$. An analysis (2001) of recent experiments
by J. Golak et al \cite{golak} and by H. Anklin et al \cite{anklin} shows
that the new data for $G_{M_{n}}/G_{D}$ points to an ondulation with
crossing point at $\sim 0.6$ (GeV/c)$^{2}$ as predicted by isospin
invariance, and Eq.(2). This ondulation was absent in the old data. A
comparison between the new data and the calculation is shown in Fig.3. For $%
Q^{2}\geq 1$ (GeV/c)$^{2}$ the calculation is in disagreement with the old
data. While the data remain close to $1$, the calculation keeps increasing.
New (unpublished) data at TJNAF appear to indicate that $G_{M_{n}}/G_{D}$
does not increase as $Q^{2}$ increases. If these data are confirmed, one
must conclude that either isospin invariance is broken above $1$ (GeV/c)$%
^{2} $ or that there are additional components in the neutron that are not
present in the proton.%
\begin{figure}
\begin{center}
\includegraphics*[width=\hsize]{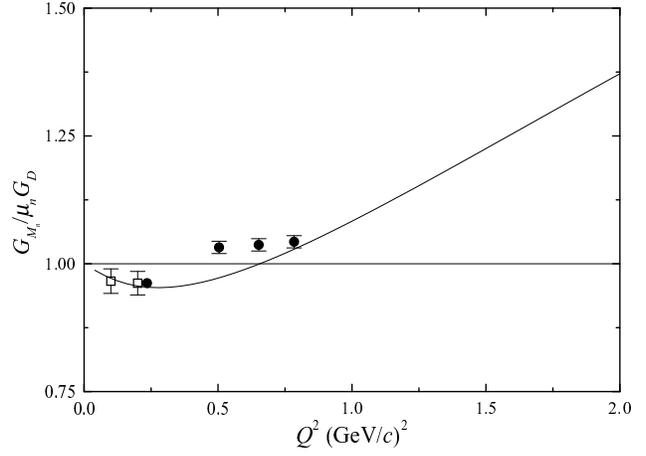}
\end{center}
\vspace{-12pt}
\caption{
Recent experimental values for $G_{M_{n}}/\mu _{n}G_{D}$ compared
with calculation. Ref. [16]: open square. Ref.[17]: filled circle.
}
\end{figure}

\subsection{The electric form factor of the neutron}

A similar situation occurs for new (1999) data for the electric form factor $%
G_{E_{n}}$ by Herberg et al \cite{herberg}, Passchier et al \cite{passchier}%
, Ostrick et al \cite{ostrick}, Rohe et al \cite{rohe}, Zhu et al \cite{zhu}%
. These are in fair agreement with the calculation as shown in Fig. 4. For $%
Q^{2}\geq 1$ (GeV/c)$^{2}$ the calculation is in disagreement with new
unpublished data. While the data remain close to $0.05$, the calculation
keeps decreasing and crosses zero at $\sim 1.4$ (GeV/c)$^{2}$. It would be
of the utmost importance to measure $G_{M_{n}}$ and $G_{E_{n}}$ at $%
Q^{2}\geq 1GeV^{2}$ in a as much as possible model independent way. A
measurement of the ratio $\mu _{n}G_{E_{n}}/G_{M_{n}}$ similar to that done
for the proton, perhaps using the reaction $d(\vec{e},e^{\prime }\vec{n})p$ 
\cite{milbrath}, will be of great value. Similar observations can be made
for $G_{E_{n}}$. In present analyses this form factor is even more sensitive
to models than $G_{M_{n}}$.%
\begin{figure}
\begin{center}
\includegraphics*[width=\hsize]{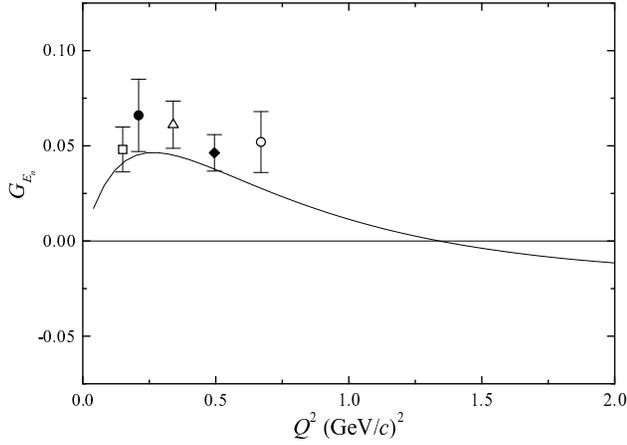}
\end{center}
\vspace{-12pt}
\caption{
Recent experimental values for $G_{E_{n}}$ compared with calculation.
Ref. [18]: open square. Ref.[19]: filled circle. Ref.[20]: filled diamond.
Ref.[21]: open up triangle.\ Ref. [22]: open circle.
}
\end{figure}

\section{SCALING LAWS}

Another important question is the extent to which the new data support
scaling laws \cite{farrar}. The parametrization of Eq.(3) is consistent with
scaling laws expected from perturbative QCD, $F_{1}\sim 1/Q^{4}$, $F_{2}\sim
1/Q^{6}$ except for $F_{2}^{V}$ whose asymptotic behavior ($Q^{2}\rightarrow
\infty $) is
\begin{equation}
F_{2}^{V}(Q^{2})\rightarrow \frac{3.706}{2\gamma ^{2}Q^{6}}\frac{m_{\rho
}^{2}+8\Gamma _{\rho }m_{\pi }/\pi }{1+\frac{\Gamma _{\rho }}{m_{\pi }}\frac{%
2}{\pi }\ln 2\sqrt{\frac{Q^{2}}{4m_{\pi }^{2}}}},
\end{equation}
that is with a weak logarithm dependence due to the effective $\rho $ mass
induced by the $\rho $ width. The scaling properties of $F_{1}$ and $F_{2}$
are determined by the only length scale in the problem, namely the size of
the intrinsic quark structure, $1/\gamma $. In order to have a quantitative
estimate of the value of $Q^{2}$ at which scaling is reached, I shall use
the following definition: a function $f(z)$ is said to be $x\%$ scaled when
its value is $x\%$ of the asymptotic value $f_{as}(z)$. The value at which
this condition is met is the solution of the equation $\mid f(z)\mid =x\mid
f_{as}(z)\mid $. For the form factors $F_{1}^{S},F_{1}^{V},F_{2}^{S}$ and
with minor modifications also for $F_{2}^{V}$, scaling properties are
determined by the function $g(Q^{2})$. Using the value $\gamma =0.25$ (GeV/c)%
$^{-2}$, one obtaines an estimate of scaling properties. The function $%
g(Q^{2})$ is $80\%$ scaled at $Q^{2}\geq 34$ (GeV/c)$^{2}$. This value is
much larger than conventionally believed, $Q^{2}\sim 4$ (GeV/c)$^{2}$. (The
dipole form $G_{D}(Q^{2})$ is $80\%$ scaled at $Q^{2}\sim 6$ (GeV/c)$^{2}$.)
The situation for the scaling properties of the form factors $G_{E}$ and $%
G_{M}$ is more complex. The parametrization of Eq.(3) is consistent, apart
from a weak logarithm dependence, with the scaling laws of perturbative QCD, 
$G_{E}\sim G_{M}\sim 1/Q^{4}$. However, relativity introduces here another
scale, $4M_{N}^{2}=3.52$ (GeV/c)$^{2}$, and, independently from the actual
value of the size scale $\gamma $, relativistic invariance requires that
scaling is not reached unless $Q^{2}$ is greater than a few times $%
4M_{N}^{2} $. (This is particularly so for the \textit{electric} form
factors). To check scaling properties it would be of utmost importance to
measure the ratio $\mu _{p}G_{E_{p}}/G_{M_{p}}$ with the recoil polarization
method beyond $10$ (GeV/c)$^{2}$.%
\begin{figure}
\begin{center}
\includegraphics*[width=\hsize]{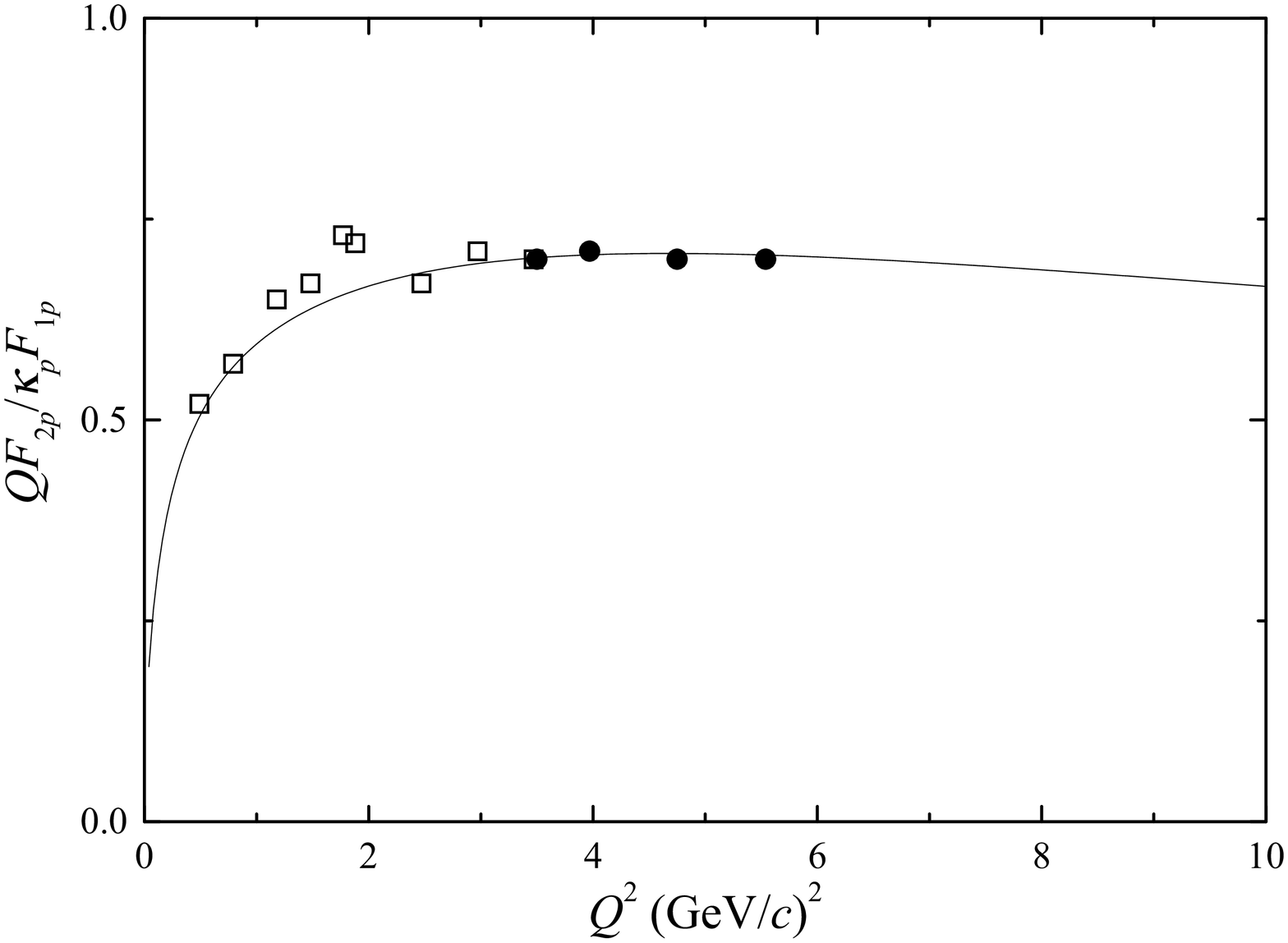}\\
\includegraphics*[width=\hsize]{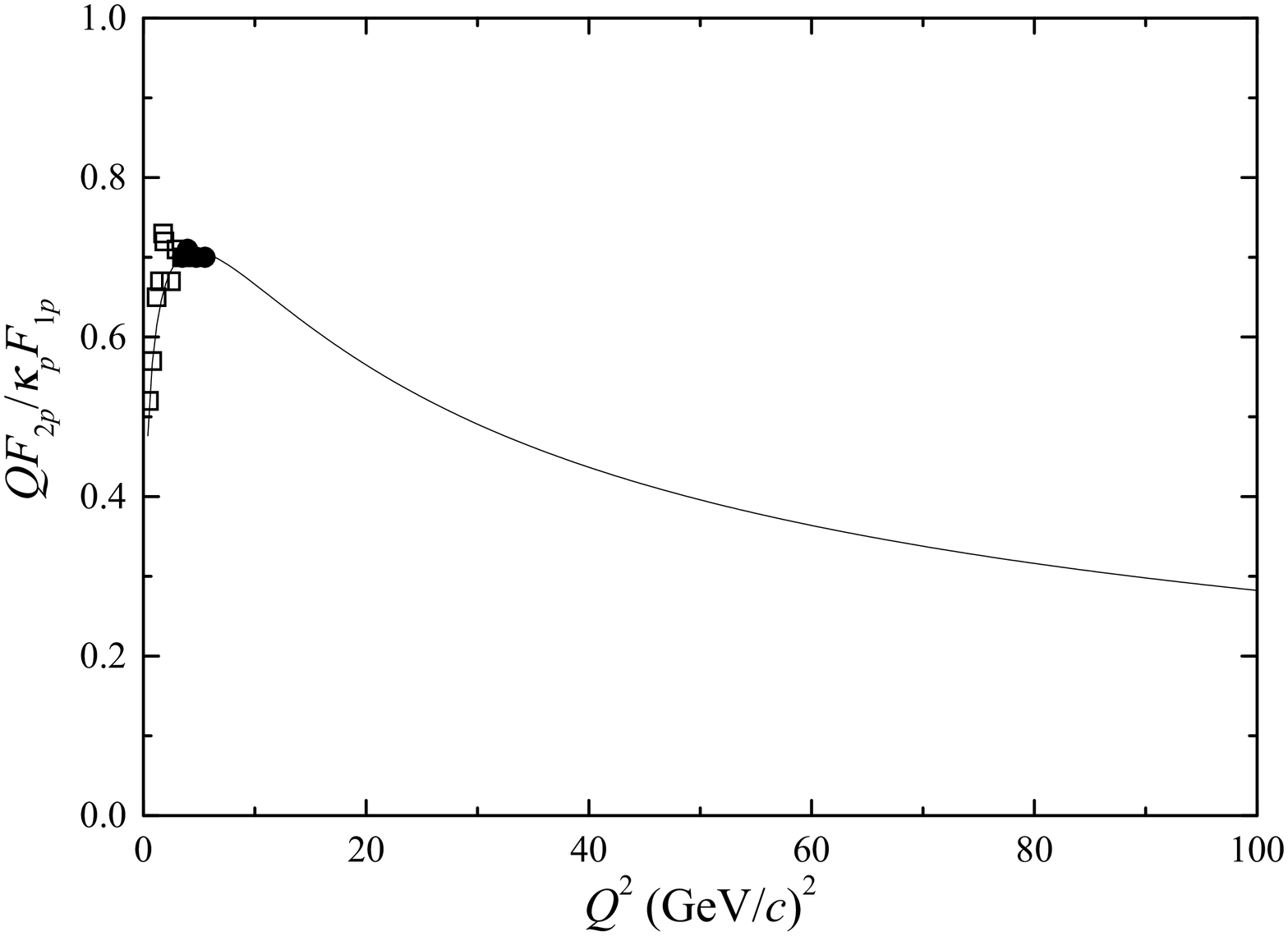}
\end{center}
\vspace{-12pt}
\caption{
The experimental ratio $QF_{2p}/F_{1p}$ compared with calculation in
the range $0\leq Q^{2}\leq 10$ (GeV/c)$^{2}$ (top) and $0\leq Q^{2}\leq 100$
(GeV/c)$^{2}$ (bottom). Ref.[2]: open square. Ref.[3]: filled circle.
}
\end{figure}

Another prediction from perturbative QCD is that the ratio $%
G_{M_{p}}/G_{M_{n}}$ approaches zero from the negative side for large $Q^{2}$,
\begin{equation}
\frac{G_{M_{p}}}{G_{M_{n}}}\rightarrow 0^{-}
\end{equation}
as a power of $\ln (Q^{2}/\Lambda ^{2})$ \cite{lepage}. The predictions of
the model discussed here are $G_{Ep}\rightarrow
-4.08/Q^{4},G_{M_{p}}\rightarrow 0.9120/Q^{4},$ and $G_{E_{n}}\rightarrow
-10.86/Q^{4},G_{M_{n}}\rightarrow -4.33/Q^{4}$ from which one can obtain
\begin{equation}
\frac{G_{M_{p}}}{G_{M_{n}}}\rightarrow -0.21.
\end{equation}
The electric values have been obtained by estimating the logarithm
dependence at $Q^{2}=100$ (GeV/c)$^{2}.$ Checking this prediction requires
the measurement of $G_{M_{n}}$ at large $Q^{2}$. Both the p-QCD result and
the 1973 result are in disagreement with the $SU(6)$ value $-3/2$ often used
in experimental analyses.

The extent to which dimensional scaling is valid has been in recent years
the subject of many investigations \cite{ralston}. It has been suggested
that the appropriate scaling variable is $QF_{2p}(Q^{2})/F_{1p}(Q^{2})$
instead of $Q^{2}F_{2p}(Q^{2})/F_{1p}(Q^{2})$. Using Eq.(3) one can easily
calculate $QF_{2p}(Q^{2})/F_{1p}(Q^{2})$. From this calculation one can see
that the quantity $QF_{2p}(Q^{2})/F_{1p}(Q^{2})$ remains flat in the
interval $2\leq Q^{2}\leq 10$ (GeV/c)$^{2}$ and drops from there on,
especially after dimensional scaling is reached at $Q^{2}\geq 34$ (GeV/c)$%
^{2}$, Fig.5. The scaling with $Q$ is thus accidental and appropriate only
to the intermediate region.

\section{STABILITY AGAINST PERTURBATIONS}

The new data clearly point out that the structure of the proton is rather
complex and that it contains at least two components. The data appear to be
in agreement in the entire measured range with a calculation in which the
two component are an intrinsic structure, presumably $q^{3}$, and a meson
cloud, $q^{3}q\bar{q}$, the latter being expressed through vector mesons ($%
\rho ,\omega ,\varphi $). The situation for the neutron is different. The
new data are in agreement with the 1973 calculation up to $1$ (GeV/c)$^{2}$.
From there on, they appear to be in disagreement with the new (unpublished)
data \cite{madey}. One can inquire whether addition of other ingredients
changes this conclusion. There are three contributions that can be analyzed
easily.

(i) The role of additional vector mesons, $\rho (1450),\omega (1390),\varphi
(1680)$ \cite{lomon}.

(ii) The addition of an intrinsic piece to the Pauli form factor
$F_{2}^{V}$. This can be done by the replacement
\begin{equation}
3.706\frac{m_{\rho }^{2}}{m_{\rho }^{2}+Q^{2}}\rightarrow (3.706-\alpha
_{\rho })\frac{1}{(1+\gamma Q^{2})}+\alpha _{\rho }\frac{m_{\rho }^{2}}{%
m_{\rho }^{2}+Q^{2}}
\end{equation}
The additional piece must be of this type to insure the proper behavior of $%
F_{2}^{V}$ for $Q^{2}\rightarrow 0$ and $Q^{2}\rightarrow \infty $.

(iii) The role of the widths of $\omega ,\varphi $ as well as the effect of
changing the width of the $\rho $ meson from the value used in \cite{IJL}.

The \textit{qualitative} features are not affected by these changes,
although quantitatively one can make some improvements on the form factor of
the neutron. However, because of isospin invariance, an improvement in the
neutron form factors produces a deterioration in the description of the
proton data. It does not appear that the problem of the neutron form factor
at large $Q^{2}$ can be solved with these changes. To solve this problem one
needs to introduce terms which act only on the neutron, that is terms with $%
F_{S}=-F_{V}$. Work in this direction is in progress.

One can also check whether the logarithm dependence of pertubative QCD
\begin{equation}
Q^{2}\rightarrow Q^{2}\frac{\ln \left[ \left( \Lambda ^{2}+Q^{2}\right)
/\Lambda _{QCD}^{2}\right] }{\ln \left[ \Lambda ^{2}/\Lambda _{QCD}^{2}%
\right] }
\end{equation}
with $\Lambda =2.27$ GeV/c and $\Lambda _{CQD}=0.29$ GeV/c \cite{gari}
produces major changes in the conclusions. This does not appear to be the
case at least up to $Q^{2}=10$ (GeV/c)$^{2}$.

\section{Time-like form factors}

By an appropriate analytic continuation in the complex plane, the form
factor of Eq.(3) can be used to analyze form factors in the time-like
region. These can be and have been experimentally obtained in the reactions $%
p\bar{p}\rightarrow e^{+}e^{-}$ and $e^{+}e^{-}\rightarrow p\bar{p}$. Two
ingredients are needed to study time-like form factors: (i) an appropriate
analytic continuation of the intrinsic form factor and (ii) an analytic
continuation of the vector meson components. For the intrinsic part, a
simple analytic continuation of the intrinsic form factor, $g(Q^{2})$, that
takes into account the complex nature of the $p\bar{p}$ interaction is
\begin{equation}
g(q^{2})=\frac{1}{(1-\gamma e^{i\theta }q^{2})^{2}}\text{ }
\end{equation}
where $q^{2}=-Q^{2}$. The parameter $\gamma $ is the same as in the
space-like region, but there is now a phase $\theta $. The width of the $%
\omega $ and $\varphi $ mesons is small and can be neglected. For the $\rho $
meson, one needs to replace \cite{frazer}
\begin{multline}
\frac{m_{\rho }^{2}}{m_{\rho }^{2}-q^{2}}\rightarrow \left(m_{\rho }+\frac{8}{\pi}\Gamma
_{\rho }m_{\pi } \right)
\Big/ \Biggl[m_{\rho }^{2}-q^{2}
\\
+\frac{(4m_{\pi }^{2}-q^{2})}{m_{\pi }}\Gamma_{\rho }\alpha (q^{2})
+i4m_{\pi }\Gamma_\rho\beta(q^{2})\Biggr]
\end{multline}
where
\begin{eqnarray}
\alpha (q^{2}) &=&\frac{2}{\pi }\left[ \frac{q^{2}-4m_{\pi }^{2}}{q^{2}}%
\right] ^{1/2}\ln \left( \frac{\sqrt{q^{2}-4m_{\pi }^{2}}+\sqrt{q^{2}}}{%
2m_{\pi }}\right)  \notag \\
\beta (q^{2}) &=&\sqrt{\frac{\left( \frac{q^{2}}{4m_{\pi }^{2}}-1\right)^3 }{%
\frac{q^{2}}{4m_{\pi }^{2}}}.}
\end{eqnarray}
The replacement (14) and (15) applies to $q^{2}\geq 4m_{\pi }^{2}$ and
should be compared with (6) and (7) that applies to $Q^{2}\geq 0$. Using the
same parameters of the 1973 calculation and adjusting the angle $\theta $,
one obtains the proton form factor $\mid G_{M_{p}}\mid $ shown in Fig. 6.
Here $\theta \simeq 53^{\circ }$. The calculation is compared with
data from \cite{castellano}, \cite{ambrogiani} \cite{bassompierre}, \cite
{bisello}, \cite{bardin}, \cite{antonelli1}, \cite{armstrong}. The large $%
q^{2}$ values \cite{armstrong}, \cite{ambrogiani}, extracted under the
assumption $\mid G_{E}\mid =\mid G_{M}\mid $, have been corrected with the
calculated $\mid G_{E}\mid $ values \cite{wan}. Apart from the threshold
behavior, presumably due to a subthreshold resonance, the overall agreement
is good. (The question of a subthreshold resonance will be discussed in a
forthcoming paper \cite{wan}.) Without further parameters one can now
calculate the neutron form factor $\mid G_{M_{n}}\mid $. A comparison with
experiment \cite{antonelli} is shown in Fig. 7. The agreement is
astonishing. In the same figures the dipole form factors, $\mu _{p}G_{D}$
and $\mu _{n}G_{D}$, are shown. One can see that the experimental data for $%
\mid G_{M_{p}}\mid $in the region $q^{2}\simeq 4-6$ (GeV/c)$^{2}$ are a
factor of 2 larger than the dipole, and those for $\mid G_{M_{n}}$ $\mid $
are a factor of 4 larger than it.%
\begin{figure}
\begin{center}
\includegraphics*[width=\hsize]{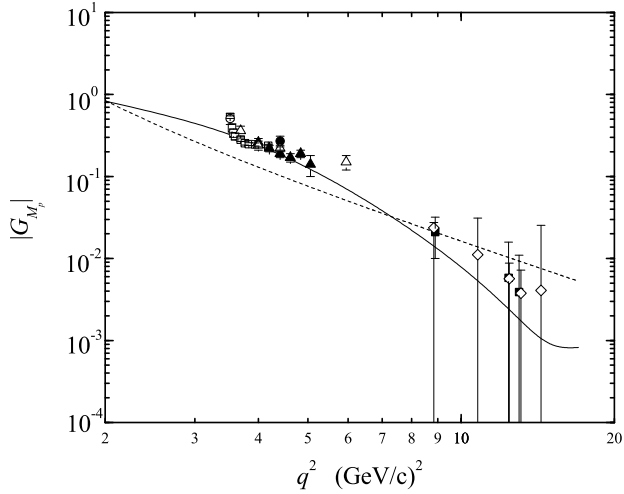}
\end{center}
\vspace{-12pt}
\caption{
Experimental values $\mid G_{M_{p}}\mid $ in the time-like region, $%
q^{2}\geq 3.52$ (GeV/c)$^{2}$, compared with calculation. Ref.[4]: filled
circle. Ref.[5]: open diamonds. Ref.[31]: open circle. Ref.[32]: filled
triangle. Ref.[33]: open square. Ref.[34]: open triangle. Ref.[35]: filled
square. The dotted line represents the dipole form factor.
}
\end{figure}
\begin{figure}
\begin{center}
\includegraphics*[width=\hsize]{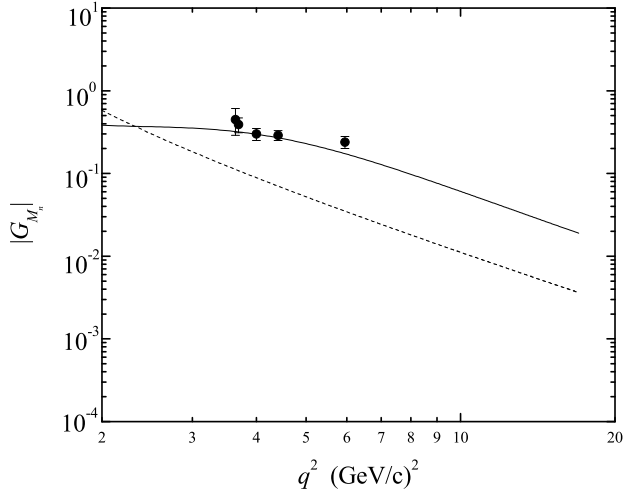}
\end{center}
\vspace{-12pt}
\caption{
Experimental values $\mid G_{M_{n}}\mid $ in the time-like region, $%
q^{2}\geq 3.52$ (GeV/c)$^{2}$, compared with calculation. Ref.[6]: filled
circle. The dotted line represents the dipole form factor.
}
\end{figure}

\section{Conclusions}

The main conclusions that one can draw from the analysis of recent
experimental data on electromagnetic form factors are:

(i) the proton appears to have a complex structure with at least two
components, an intrinsic component (valence quarks) and a meson cloud ($q%
\bar{q}$ pairs). The size of the intrinsic structure is r.m.s. $\sim 0.34$
fm.

(ii) Perturbative QCD is not reached in the proton up to $Q^{2}\sim 10$
(GeV/c)$^{2}.$ Physics up to this scale is dominated by a mixture of
hadronic and quark components.

(iii) Symmetry (in particular relativistic invariance), rather than detailed
dynamics, appears to be the determining factor in the structure of the
proton.

The situation appears to be different for the neutron. Here recent
experimental data up to $1$ (GeV/c)$^{2}$ are consistent with isospin
invariance and the structure of the proton, while preliminary data at $%
Q^{2}\geq 1$ (GeV/c)$^{2}$ appear to indicate that either isospin invariance
is broken or that additional components play a role. It would be of the
utmost importance to understand this discrepancy.

An analysis of the the form factors in the time-like region indicates that
both proton and neutron data are consistent with the two component picture
introduced in 1973, in particular the oberved ratio $\mid
G_{M_{n}}/G_{M_{p}}\mid \sim 2$, appears to be a consequence of this
structure and is easily explained. In order to understand this basic point
of the structure of the nucleon, it would be of utmost importance to
remeasure $G_{M_{n}}$ with better accuracy, possibly extending the
measurement to larger $Q^{2}$ values.

\section{Aknowledgements}

This work was performed in part under DOE Grant No. DE-FG-02-91ER40608. I
wish to thank Rinaldo Baldini for bringing to my attention the neutron
time-like data used in Fig.7.

\end{document}